\documentclass[12pt]{article}

\usepackage{amssymb,epsfig}

\begin{document}

\begin{center}

{\Large \bf Simulations of muon-induced neutron flux at large depths
underground}

\vspace{0.5cm}

{\large V. A. Kudryavtsev~\footnote{Corresponding author, 
e-mail: v.kudryavtsev@sheffield.ac.uk}, N. J. C. Spooner,
J. E. McMillan}

\vspace{0.3cm}
{\it Department of Physics and Astronomy, 
University of Sheffield, Sheffield S3 7RH, UK}

\vspace{1cm}

\begin{abstract}
The production of neutrons by cosmic-ray muons at large depths 
underground is discussed. The most recent versions of the muon propagation 
code MUSIC, and particle transport code FLUKA are used to 
evaluate muon and neutron fluxes. The results of simulations are
compared with experimental data.
\end{abstract}

\end{center}

\vspace{0.2cm}


\vspace{0.2cm}
\noindent Corresponding author: V. A. Kudryavtsev, Department of Physics 
and Astronomy, University of Sheffield, Hicks Building, Hounsfield Rd., 
Sheffield S3 7RH, UK

\noindent Tel: +44 (0)114 2224531; \hspace{2cm} Fax: +44 (0)114 2728079; 

\noindent E-mail: v.kudryavtsev@sheffield.ac.uk
\pagebreak

{\large \bf 1. Introduction}
\vspace{0.3cm}

\indent Existing and planned dark matter searches and 
neutrino detection experiments require very sensitive equipment and 
sophisticated 
data acquisition capable of tagging events from the source and 
discriminating them from all kinds of background. Some background, 
however, is undistinguishable from the expected signal events. As an 
example of this we can mention neutron background in experiments 
searching for dark matter particles in the form of Weakly Interacting 
Massive Particles (WIMPs), also known as neutralinos, 
the lightest supersymmetric 
particle in SUSY models. WIMPs are expected to interact with ordinary 
matter in detectors to produce nuclear recoils, which can be 
detected through ionisation, scintillation or phonons. Identical 
events can be induced by neutrons. Thus, only suppression of any background
neutron 
flux by passive or active shielding will allow experiments to 
reach sufficiently high sensitivity to neutralinos. Designing shielding for such 
detectors requires simulation of neutron fluxes from various sources.

Neutrons underground arise from two sources: i) local 
radioactivity, and ii) cosmic-ray muons. Neutrons associated with 
local radioactivity are produced mainly via ($\alpha$, n) reactions, 
initiated by $\alpha$-particles from U/Th traces in the rock and 
detector elements. Neutrons from spontaneous fission of $^{238}$U 
contribute also to the flux at low energies. The neutron yield from 
cosmic-ray muons depends strongly on the depth of the underground 
laboratory. It is obvious that suppression of the muon flux by a large 
thickness of rock will also reduce the neutron yield. The dependence, 
however, is not linear. 

In general, at large depth underground the 
neutron production rate due to muons is about 3 orders of magnitude less 
than that of neutrons arising from local radioactivity. 
(Note that this figure depends strongly 
on the depth and U/Th contamination). 
The muon-induced neutron flux can
be, however, important for experiments intending to reach high 
sensitivity to WIMPs or to low-energy neutrino fluxes. There are 
several reasons for this: 1) the energy spectrum of muon-induced neutrons 
is hard, extending to GeV energies, and fast neutrons can travel far from 
the associated
muon track, reaching a detector from large distances; 2) fast neutrons 
transfer larger energies to nuclear recoils making them visible in 
dark matter detectors, while many recoils from $\alpha$-induced neutrons
fall below detector energy thresholds; 3) a detector 
can be protected against neutrons from the rock activity by 
hydrogen-rich material, possibly with addition of thermal neutron 
absorber; such a material, however, will be a target for cosmic-ray 
muons and will not protect against muon-induced neutrons. The only way 
to reduce this flux is to add an active veto, rejecting all events 
associated with passing muons.

In this paper we discuss neutron production by muons at large 
depth underground. There are several reactions which lead to  
neutron appearance. The main processes are: 
a) negative muon capture; this process occurs with 
stopping muons and plays a significant role only at shallow depth; b) 
muon-induced spallation reactions; c) neutron production by hadrons 
(and photons) in muon-induced (via photonuclear interaction) 
hadronic cascades; d) neutron production by photons in 
electromagnetic cascades initiated by muons.

The muon-induced neutron case was first discussed by Zatsepin and 
Ryazhskaya \cite{zr} and its importance for large proton decay and 
neutrino experiments was studied \cite{russ1}. Some results are 
summarised in Ref. \cite{russ2}. Several 
measurements of the neutron flux underground have been performed 
\cite{bezrukov,asd,lsd,lvd,paloverde,chen}. However, no precise 
three-dimensional simulation of neutron production has been done for 
most experiments. Recently accurate versions of Monte Carlo codes FLUKA 
\cite{fluka} and GEANT4 \cite{geant4}, designed for particle 
transport over a wide energy range, have become available. They are capable 
of simulating neutron production by muons, hadrons and photons,
as well as neutron transport and interactions in the detector volume.
Simulations of neutron production by muons with FLUKA have been 
performed by Battistoni et al. \cite{battistoni} and 
Wang et al. \cite{wang}. The authors of Ref. \cite{wang} studied neutron 
production by muons in scintillator and compared the results with 
existing data. Recently extensive neutron background studies have
been initiated by the joint EDELWEISS/CRESST team of dark matter
experiments at Gran Sasso and Modane \cite{chardin}.

In this work we have used the FLUKA code for neutron production and 
transport. We have calculated the neutron production rate in 
scintillator for several muon energies and 
have compared our results with those from Ref. \cite{wang} and with
experimental data.
We have studied the neutron production rate as a function of the atomic 
weight of the target material. We have calculated the neutron flux as a 
function of distance from muon tracks, which is very important for 
designing muon veto systems, 
and we have compared the result with available data. 
Finally, we have used MUSIC \cite{music,vak1,vak2} to simulate the
muon energy spectrum underground and we have calculated neutron 
production from the muon spectrum. 

The work has been done as a part of a programme of
neutron background studies for the dark matter experiments at Boulby 
mine (North Yorkshire, UK) (see Ref. \cite{boulby} for a review of
dark matter searches at Boulby). We hope that other experiments having 
similar background problems will benefit from this work too.

The paper is organised in the following way. Muon fluxes, energy 
spectra and their uncertainties are discussed in Section 2. 
Simulations of muon-induced neutrons are presented in Section 3. The
conclusions are given in Section 4.

\vspace{0.5cm}
{\large \bf 2. Muon flux and energy spectrum underground}
\vspace{0.3cm}

Knowledge of the muon flux is clearly 
important for calculations of the neutron flux or 
counting rate -- it provides the absolute normalisation.
The muon energy spectrum at a given experimental site also affects the neutron 
production rate. Thus precise 
knowledge of the muon spectrum and absolute normalisation is crucial for 
neutron flux simulations.

To calculate the muon fluxes and energy spectra at large depths 
underground, we used here parameterisation of the surface muon spectrum
propagated through the rock using the MUSIC code. The calculation was
performed in the following way. First, muons with various 
energies at the surface were transported using MUSIC 
\cite{music,vak1} down to 15 km w.e. of rock underground and their energy 
distributions at various depths stored on disk. For a set of muon energies 
at sea level, and for a set of different depths underground, 
the muon energy distributions 
were obtained in the form
$P(E_{\mu},X,E_{\mu 0})$ -- the probability 
for a muon with energy $E_{\mu 0}$ at the surface to have energy $E_{\mu}$ at 
depth $X$. If no muon with a certain energy survives, 
then the probability to 
reach the depth will be equal to 0. Muons were propagated in several 
types of rock, including standard rock, Gran Sasso rock and Boulby 
rock.

Special codes were used to calculate differential and integral 
muon intensities at the various depths (code SIAM) and to simulate muon 
spectra and angular distributions (code MUSUN). The codes had 
previously been used to simulate single atmospheric muons under water 
\cite{vak2}. A brief description of the codes \cite{vak3} is given below.

To calculate the differential muon intensity underground the 
following equation was used:

\begin{equation}
I_{\mu}(E_{\mu},X,\cos\theta)=\int_0^{\infty} P(E_{\mu},X,E_{\mu 0})
{{d I_{\mu 0}(E_{\mu 0},\cos\theta^{\star})} \over {d E_{\mu 0}}}
d E_{\mu 0}
\label{muon intensity}
\end{equation}

\noindent where 
${{d I_{\mu 0}(E_{\mu 0},\cos\theta^{\star})} \over {d E_{\mu 0}}}$
is the muon spectrum at sea level at zenith angle
$\theta^{\star}$ (the zenith angle at the surface, $\theta^{\star}$,
was calculated from that 
underground, $\theta$, taking into account the curvature of the Earth). 

The energy spectrum at sea level was taken either according to the 
parameterisation proposed by Gaisser \cite{gaisser} (modified 
for large zenith angles \cite{lvd1}) or following the best fit to the 
`depth -- vertical muon intensity' relation measured by the LVD 
experiment \cite{lvd1}. The first parameterisation \cite{gaisser} has 
a power index for the primary all-nucleon spectrum of 2.70, while the 
second one \cite{lvd1} uses the index 2.77 with normalisation 
to the absolute flux measured by LVD. 
The difference between the results obtained 
with these two spectra shows a possible spread in the 
muon energy spectra at the surface and, hence, in muon 
intensities underground.

The ratio of prompt muons (from charmed particle decay) 
to pions was chosen as 
$10^{-4}$, which was well below an upper limit set by the LVD 
experiment \cite{lvd2}. Note, however, that the prompt muon flux does 
not significantly affect muon intensities even at large depths.

To calculate the integral muon intensity, an integration of 
$I_{\mu}(E_{\mu},X,\cos\theta)$ over $dE_{\mu}$ was carried out. 
An additional integration over $\cos\theta$ defined the global intensity
for a spherical detector.

The muon energy spectrum, fraction of prompt muons, type of 
rock and depth can be chosen in the SIAM and MUSUN codes.

The muon intensities, calculated with SIAM for two types of 
rock, two parameterisations of muon energy spectrum at the surface and 
at several depths 
are given in Table \ref{intensity}. Our present calculations agree well with 
previous results obtained with MUSIC \cite{music}. 
Note that only a few modifications (including more recent 
cross-sections) have been implemented in the code \cite{vak1} since the first 
version was published \cite{music}.

The absolute muon flux (intensity) underground depends on the 
surface relief, which should be taken into account for 
any particular experiment. For a complex mountain profile, as for the 
Gran Sasso and Modane underground laboratories, it is difficult to 
give predictions of the muon flux without precise knowledge of the slant 
depth distribution. For a flat surface above a detector the 
calculations are straightforward. We present here the results for 
vertical and global (integrated over solid angle for 
spherical detector) muon intensities under a flat surface for standard 
rock ($<Z>=11, <A>=22$). We used parameterisation of the muon 
spectrum at sea level according to 
a best fit to the `depth -- vertical intensity' relation measured 
by the LVD experiment at Gran Sasso. Note that the LVD results agree 
well with those of the MACRO experiment \cite{macro}.

The ratio of global intensity (column 3) to the vertical one (column 2) 
gives an average solid angle for a
particular depth under a flat surface. It decreases from about 2 sr 
down to about 0.5 sr for depths from 0.5 km w.e. down to 10 km w.e., 
being about 1 sr at 3-4 km w.e., at which many experiments are located.

Intensities calculated with Gaisser's parameterisation of muons 
at the surface \cite{gaisser} (column 4) 
are lower by 20\% at small depths
and higher by 15\% at large depths compared to the LVD 
parameterisation (column 3). This is due to differences in the 
power index of the muon spectrum and absolute normalisation. Gaisser's 
parameterisation describes reasonably well the muon data at low energies 
(below 1 TeV), while the LVD data provide a good parameterisation for the muon 
spectrum above 1.5-2 TeV (for depths more than 3 km w. e.).

The muon intensities in Boulby rock (column 5), which has slightly 
higher mean values of atomic number and weight 
($<Z> \approx 11.7, <A> \approx 23.6$), are smaller than in 
standard rock (column 3) due to the larger muon energy losses, 
which are proportional 
to $Z^{2}/A$. The difference, however, does not exceed 5\% at depths 
smaller than 2 km w.e., where the ionisation energy loss (proportional 
to $Z/A$) dominates. At 3 km w.e. the difference is about 8\%, which 
gives an uncertainty in $<Z>$ and $<A>$ of about 2-3\% acceptable from 
the point of view of the accuracy required for muon and neutron flux 
simulations here.

The muon energy spectrum underground is also important for neutron flux 
simulations, since neutron production rates increase with 
muon energy. While the absolute muon flux underground can be measured to 
check simulations, the muon energy spectrum is hard to determine 
experimentally. We have to rely on simulations. The mean muon energy 
is a parameter, which characterises well the muon energy spectrum. 
Table \ref{energy}  shows the calculations of mean muon energies underground.

An important conclusion, which can be derived from comparison of the
mean muon energies for global fluxes, is that their spread is much 
smaller than that for the fluxes themselves. The mean muon energies 
are very similar for different types of rock at small depths and 
differ by no more than 5\% at large depths. The two different 
parameterisations of the muon spectrum at sea level result in the 
$\approx 5\%$ difference in mean energy. For a fixed slant depth 
the mean energy for inclined muon directions ($60^{\circ}$ 
in our simulations) is very close to that at vertical. The increase 
in mean energy for the global flux compared to the vertical flux is due to 
the increase of slant depth with zenith angle. With 
increasing depth, the effective solid angle decreases 
(see Table \ref{intensity} 
and the discussion above), and the mean muon energy for global flux 
becomes closer to that at vertical.

It is interesting to estimate uncertainties in the muon flux and mean 
energy for a fixed depth due to uncertainties in depth, density, 
rock composition and parameterisation of the muon spectrum at the surface.  
Depth and rock density have a similar effect on the muon flux, 
since the intensity depends on their product -- 
column density expressed in metres of water equivalent or hg/cm$^{2}$.
Changing the column density (either depth or density) by 2\% results 
in a 10\% change in muon flux at 3 km w.e. and has no effect on the 
muon energy spectrum. An increase of $<Z>$ and $<A>$ by about 
7\% (changing from standard to Boulby rock, for example) 
without changing the 
density gives an 8\% decrease in the flux and a 3\% increase in the mean 
energy at 3 km w.e. 
Finally, 7\% uncertainty in the flux and 4\% uncertainty in 
the mean energy at 3 km w.e. may arise from the difference in  
parameterisations of the muon spectrum at sea level. Note, however, that 
the LVD experiment provides direct measurements of muon intensities 
at 3 km w.e. and below, which were used to derive the 
parameterisation for muon flux at sea level. So, unless the 
experimental site is at shallow depth, the LVD parameterisation is 
the preferred option.

Two measurements of mean muon energy underground carried out with
the NUSEX \cite{nusex} ($<E_{\mu}>=346 \pm 14 \pm 17$ GeV for vertical muons
at 5 km w.e. in standard rock) and MACRO \cite{macro1} 
($<E_{\mu}>=270 \pm 3 \pm 18$ GeV for single muons
at 3.0-6.5 km w.e. in standard rock) detectors are in reasonable
agreement with our simulations (see Table \ref{energy}).
Note that the experimental errors quoted by the authors \cite{nusex,macro1}
are larger than the possible systematic uncertainty of the
simulations with a particular code. There is a large spread
of muon intensities and mean energies as calculated with a number of
muon propagation codes (see, for example discussion in Ref. 
\cite{nusex}). Note, however, that in our simulations 
we used the results obtained with the MUSIC code, which was tested
against experimental underground muon data \cite{lvd1,lvd2,sno}.

\vspace{0.5cm}
{\large \bf 3. Simulations of muon-induced neutrons}
\vspace{0.3cm}

As can be seen from Tables \ref{intensity} and \ref{energy} and discussion 
above, 
the uncertainty in absolute muon flux is larger than that of the 
muon energy spectrum underground (mean energy). This conclusion is 
encouraging because the absolute flux can be measured directly, while it
is difficult to determine the mean muon energy from experiment.
Thus, the uncertainty in neutron flux is almost directly proportional 
to that in the muon flux.

Neither code provides an absolute accuracy in the neutron simulation.
There is always an uncertainty related to our knowledge of the
neutron production mechanism and to the choice of the model for its description.
In our opinion the FLUKA \cite{fluka} and GEANT4 \cite{geant4} codes are
the best suited for this job.
We simulated neutron production by muons with FLUKA 
\cite{fluka}. The models of photoproduction and hadronic interactions, 
used in FLUKA, are described in the Refs. \cite{fluka,wang}.

It is widely accepted (see, for example, Ref. \cite{russ2,lsd,wang} and 
references therein) that the neutron production at a certain depth can be 
approximated by assuming that neutrons are produced by muons, 
all having mean energy corresponding to this depth. 
Keeping this in mind we started with 
the simulations of neutron production in scintillator (C$_{10}$H$_{20}$)
as a function of muon energy in order to compare the results with available 
experimental data and previous simulations with FLUKA \cite{wang}.

As many neutrons are produced in large cascades initiated by muons, the 
equilibrium between neutron and muon fluxes 
(when the ratio of neutron to muon fluxes is constant)
begins only when a muon has crossed 
a certain thickness of a medium. This is because cascades 
need some depth to develop and produce neutrons. So, the thickness of 
medium was chosen large enough (of the order of 4000-5000 g/cm$^2$)
for such an equilibrium to take 
place, and only neutrons in a reduced layer of a medium (where the 
equilibrium is in place) were counted. 

Large thicknesses of material produce, however, another effect: the muon 
energy can be reduced compared to the initial value due to 
interactions with matter. This is particularly important for low 
energies. We checked carefully that the neutron flux in our simulations
did not decrease with the thickness of a medium crossed by muons. 
Where a reduction in flux 
could not be avoided (for low muon energies), the appropriate 
correction of the order of a few percent was applied.

The FLUKA code returned the number of neutrons in various layers in a 
medium and neutron spectra in various regions and at the boundaries 
between regions. Some neutrons, however, are counted twice (see also 
the discussion in Ref. \cite{wang}). This happens when a neutron 
produces a star. Then, the scattered neutron is counted also as a 
secondary neutron (with different energy). To avoid double counting, 
the number of stars produced by neutrons was subtracted from the 
total number of neutrons.

\begin{figure}[htb]
\begin{center}
\epsfig{figure=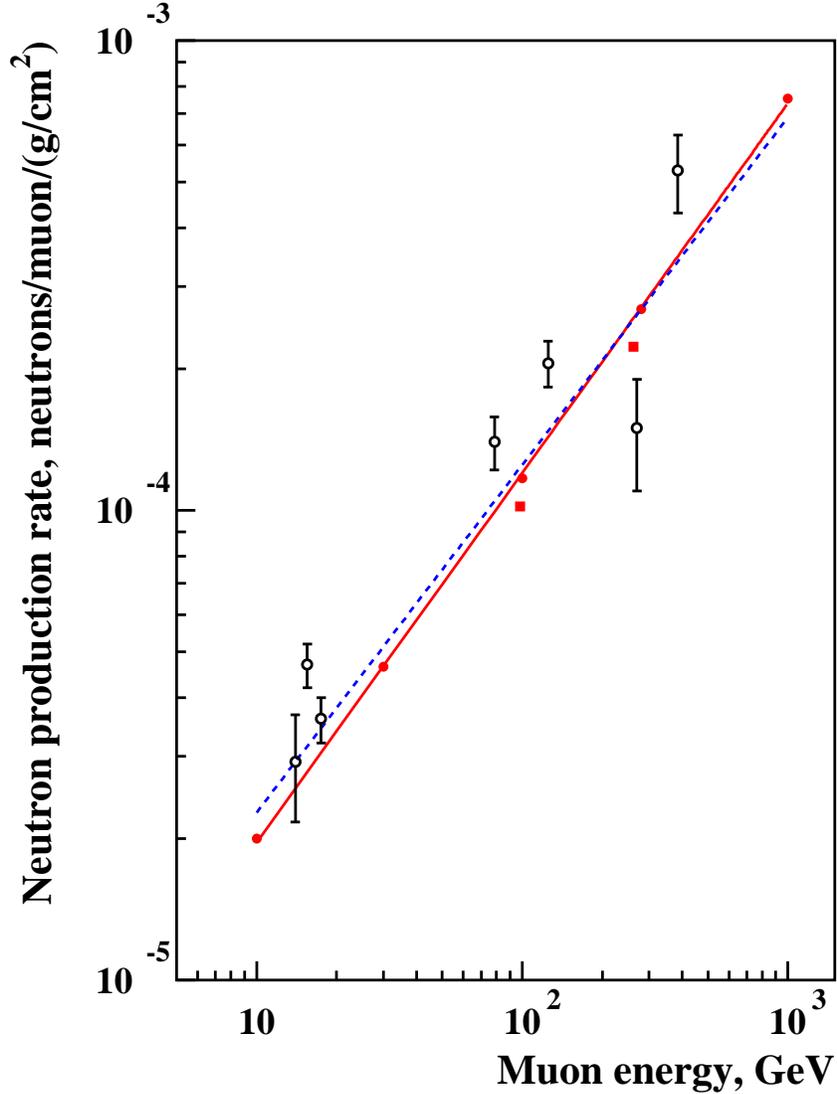,height=15cm}
\caption{Average number of neutrons produced by a muon per unit path 
length (1 g/cm$^{2}$) in scintillator as a function of muon energy. Our 
results are shown by filled circles. The parameterisation with Eq. 
(1) is shown as a solid line. The parameterisation found in Ref.
\cite{wang} is plotted by a dashed line. The measurements shown are as follows
(in order of increasing energy): 20 m w.e. (minimal depth) 
\cite{chen,paloverde}, 25 m w.e. \cite{bezrukov}, 
32 m w.e. \cite{paloverde}, 316 m w.e. 
\cite{bezrukov}, 570 m w.e. \cite{asd}, 3000 m w.e. \cite{lvd}, 
5200 m w.e. \cite{lsd}. Filled squares show the number of neutrons, produced
in scintillator by muons with a real spectrum for depths of 0.55 km w.e. 
and 3 km w.e. in Boulby rock (mean energies 98 and 264 GeV, respectively).} 
\label{muenergy}
\end{center}
\end{figure}

The average number of neutrons produced by a muon per unit path length 
(1 g/cm$^{2}$)
in scintillator is presented in Figure \ref{muenergy}
as a function of muon energy. 
Our results (filled circles) have been fitted to a function:

\begin{equation}
R_{n} = a \times E^{\alpha}
\label{muen}
\end{equation}

\noindent where $a=(3.20 \pm 0.10) \times 10^{-6}$ and $\alpha=0.79 \pm 0.01$.
The agreement with simulations by Wang et al. \cite{wang} (dashed 
line shows the fit) is pretty good. Note that Wang et al. 
\cite{wang} performed the simulations in scintillator with a slightly 
different fraction of carbon and hydrogen atoms (C$_{10}$H$_{22}$), 
but this should not significantly affect the neutron production. The small 
difference between the simulation results can be attributed to a 
number of corrections described above and in Ref. \cite{wang}. 

Also shown in Figure \ref{muenergy} 
are the measurements of neutron production by 
several experiments. In the experiments neutrons were produced by 
muons with a certain spectrum. Here we have plotted their results as a 
function of mean muon energy underground. For experiments at large 
depths (more than 400 m w.e.) we used the mean 
muon energy calculated by the authors. Our simulations, however, 
predict smaller mean muon energies at these depths with 
the exception of the LVD result (270 GeV). 
(Note that, to evaluate mean muon 
energy for the LVD experiment, the same MUSIC propagation code and 
LVD depth--intensity curve were used by the authors 
\cite{lvd1}). For shallow depths we used estimates of the
mean muon energies from Ref. \cite{wang}. 

All measurements, except those of LVD, show higher neutron production rate 
than simulations with FLUKA. Similar conclusion was reached in 
Ref. \cite{wang}. If the data points were shifted to smaller muon 
energies, then the disagreement would become even more prominent. So, 
an error in the calculation of mean muon energy cannot explain the 
difference.

Another possible explanation could be the difference in neutron 
production between muons with fixed mean energy and muons with a real 
spectrum underground. 
This was checked by simulating neutron production in scintillator by 
muons with a real spectrum for depths of 0.55 km w.e. and 3 km w.e. 
in Boulby rock (mean energies 98 and 264 GeV, respectively; filled 
squares in Figure \ref{muenergy}). In both 
cases a smaller neutron production rate in scintillator was found. 
The difference is of the order of (10-15)\% for large range of depths 
(0.5 -- 3 km w.e.) and mean muon energies (100 -- 300 GeV).
(Note that a few percent decrease in the neutron production is expected
due to the attenuation of the muon flux with realistic spectrum with
mean energy of 280 GeV when muons are crossing a few tens of m w.e.).
A similar difference was found also 
for neutron production in NaCl salt. For marl rock (mainly
CaCO$_3$), however, the difference is not significant, the neutron
production rate for mean muon energy of 280 GeV being about 
$4.0 \times 10^{-4}$ neutrons/muon/(g/cm$^3$). Neutron
production rate in lead by muons with an energy spectrum with
mean energy of 280 GeV (as at 3.2 km w.e. depth in standard rock) 
is 10\% higher than that by muons with a fixed energy of 280 GeV.
The difference between neutron production by a muon spectrum and by muons
with fixed mean energy and the dependence of this effect on the target
material emphasise the importance of the simulations with realistic input
data, such as muon energy spectrum and target composition.
The smaller neutron production rate in scintillator calculated with a real 
muon spectrum makes agreement 
between LVD data and simulations better, while other experimental 
results are less consistent with predictions. 

The difference between LSD and LVD measurements (385 GeV and 270 GeV, 
respectively) represents a real puzzle. Although performed at 
different depths, the experiments used a similar modular structure, 
similar liquid scintillator counters and similar analysis techniques.
Clearly, better study of systematic effects is needed.

In our simulations with fixed muon 
energies we did not observe any significant difference in 
neutron production by positive and negative muons with energies above 
10 GeV, which proved the absence of muon capture. 
Only the real muon spectrum at small depth could, in principle, produce 
a significant number of stopping muons and, consequently, neutrons 
through negative muon capture. At large depth, where the number of stopping
muons is negligible, negative muon capture does not contribute much to 
neutron production.

\begin{figure}[htb]
\begin{center}
\epsfig{figure=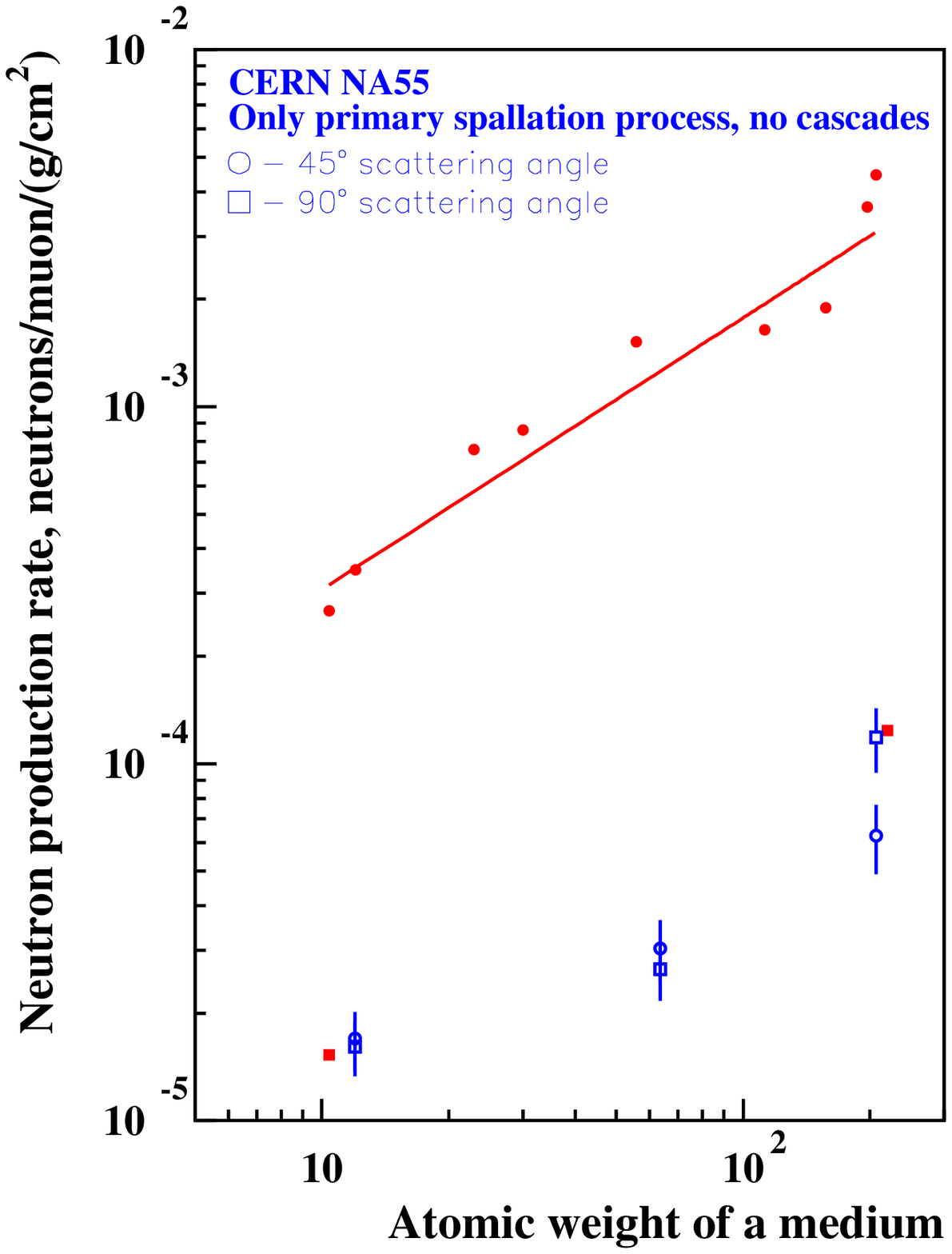,height=15cm}
\caption{Dependence of neutron rate on the 
atomic weight of material. Materials and compounds used in the 
simulations and presented in the figure by filled circles
are (in order of increasing atomic weight):
C$_{10}$H$_{20}$ ($<$$A$$>$=10.4), C ($<$$A$$>$=12.0), 
Na ($<$$A$$>$=23.0), NaCl ($<$$A$$>$=30.0), 
Fe ($<$$A$$>$=55.9), Cd ($<$$A$$>$=112.4), 
Gd ($<$$A$$>$=157.3), Au ($<$$A$$>$=197.0), and Pb ($<$$A$$>$=207.2). 
A simple 
parameterisation by a power-law is given by the solid line. Also shown 
are the measurements of neutron rate by NA55 at CERN in 
three thin targets at two neutron scattering angles (open circles and 
open squares) (see text for details). These are normalised at small atomic 
weight (carbon) to our simulations 
with FLUKA for the muon spallation only (filled squares). 
For the muon spallation our result for lead is artificially shifted to
A=220 to avoid the overlaping with one of the CERN points.}
\label{n-a}
\end{center}
\end{figure}

We studied also the dependence of neutron rate on the 
atomic weight of material. The neutron rate was obtained with 
280 GeV muon flux in several materials and compounds and is shown in 
Figure \ref{n-a} by filled circles. The errors of the simulations do not 
exceed 5\% and are comparable to the size of the circles on the 
figure. It is obvious that on average 
the neutron rate increases with 
the atomic weight of material, but no exact parameterisation was 
found which would explain the behaviour for all elements and/or 
compounds. The general trend can be fitted by a simple power-law 
form (solid line in Figure \ref{n-a}):

\begin{equation}
R_{n} = b \times A^{\beta}
\label{natomic}
\end{equation}

\noindent where $b=(5.33 \pm 0.17) \times 10^{-5}$, $A$
is the atomic weight (or mean 
atomic weight in the case of a compound) and $\beta=0.76 \pm 0.01$.
It is clear from the figure, however, that the points on 
Figure \ref{n-a} are largely spread around the line. We compared our
results with measurements performed in the NA55 experiment at CERN 
with a 190 GeV muon beam \cite{na}. 
The neutron production was measured in thin 
targets at several neutron scattering angles, so direct comparison 
with our simulations is difficult. The use of thin targets 
allowed the measurement of neutron production in the first muon 
interaction only (without accounting for neutrons produced in cascades)
\cite{na}. We calculated neutron production in the first muon 
interaction in scintillator and lead and plotted it in Figure 
\ref{n-a} (filled squares) together with the measurements \cite{na} 
at two scattering angles (open circles and open squares). Since the 
measured values refer to particular scattering angles, we normalised 
them to our results at small atomic weight (carbon). The measured 
behaviour of the neutron rate with atomic weight agrees well with 
FLUKA predictions.

The processes contributing to neutron flux were studied in Ref. 
\cite{wang} for a liquid 
scintillator as a target. We performed a similar 
investigation for scintillator and extended it to heavy targets. We 
subdivided the neutron production by muons into three main processes: 
i) direct muon-induced spallation (first muon interaction), ii) 
muon-induced hadronic cascades, and iii) muon-induced electromagnetic 
cascades. We neglected negative muon capture, which contributes only 
at small depths or small muon energies. We found that in scintillator 
at a muon energy of 280 GeV, 75\% of neutrons are produced in hadronic 
cascades, 20\% in electromagnetic cascades, 
and 5\% in muon-induced spallation. The last number will be increased
up to 9\% if secondary neutrons produced in collisions of neutrons from
muon spallation with nuclei are counted here and not in the hadronic cascades. 
For 10 GeV muons the contribution from 
hadronic cascades drops to 38\%, while electromagnetic cascades 
contribute to 35\% of neutrons, and muon spallation is 
responsible for 27\% of neutrons (36\% if secondary neutrons are counted here). 
This is due to a smaller muon photoproduction cross-section at these energies. 
(Note that the error in the calculations at low energies is quite large, 
about 3-5\%, due to the small number of neutrons produced and large 
corrections involved).

For a heavy target (e.g. lead) the contribution from electromagnetic cascades 
becomes more important (42\% for 280 GeV muons) 
because the cross-section of electromagnetic muon interactions 
is proportional to $Z^{2}/A$. Hadronic cascades give about 55\% of 
neutrons for 280 GeV muons and muon-induced 
spallation is responsible for the remaining 3\% (10\% if secondary
neutrons from neutron-nucleus collisions are counted here and not in hadronic
cascades).

\begin{figure}[htb]
\begin{center}
\epsfig{figure=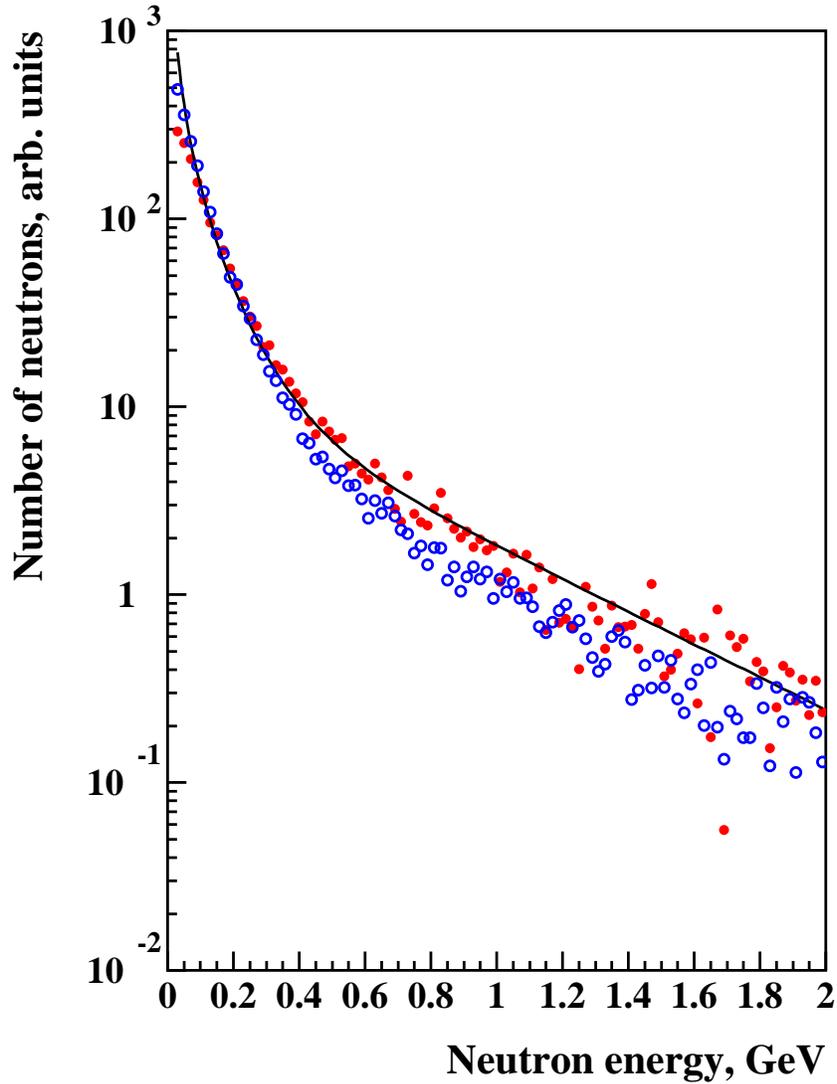,height=15cm}
\caption{Neutron energy spectrum in scintillator (filled circles) 
and NaCl (open circles) obtained with muon spectrum at about 3 km w.e. 
underground. Parameterisation proposed
in Ref. \cite{wang} for scintillator for muon energy 280 GeV 
is shown by the solid line, 
arbitrarily normalised to our simulations to reach visual agreement.
Arbitrary units are used for all spectra, the normalisation being 
provided by the total neutron production rate (see Figures 
\ref{muenergy} and \ref{n-a}).}
\label{nspectrum}
\end{center}
\end{figure}

The neutron energy spectrum was calculated for various targets. 
Figure \ref{nspectrum} shows the spectrum obtained for scintillator 
and NaCl together with parameterisation proposed for scintillator in 
Ref. \cite{wang}. In our simulations the real muon spectrum at about 3 km 
w. e. underground was used. The parameterisation for scintillator 
\cite{wang} was obtained for 280 GeV muons, which is close to the 
mean energy of the muon spectrum used in the present work. Two 
conclusions can be derived from figure \ref{nspectrum}: i) the 
parameterisation proposed in Ref. \cite{wang} agrees with our 
simulations only at neutron energies higher than 50 MeV; this may be 
partly due to the fact that in our simulations all muon energies 
contribute to the neutron production, ii) the neutron 
energy spectrum becomes softer with increase of $<A>$, although 
the total neutron production rate increases (see Figure \ref{n-a}). 

\begin{figure}[htb]
\begin{center}
\epsfig{figure=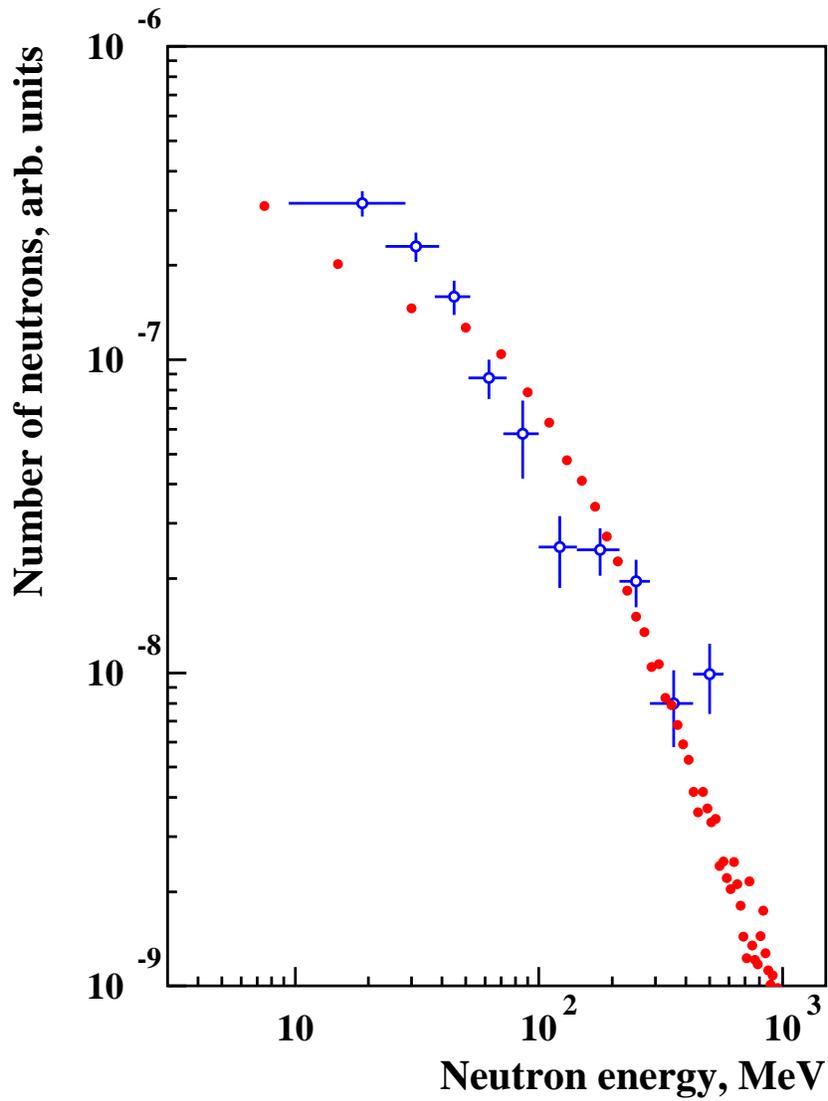,height=15cm}
\caption{Neutron energy spectrum in scintillator (filled circles) 
in comparison with the LVD data \cite{lvd}. LVD data are
normalised to the calculated spectrum to reach better visual agreement. 
The absolute normalisation is provided by the total neutron
production rate (see Figure \ref{muenergy}). }
\label{nspectrum1}
\end{center}
\end{figure}

Figure \ref{nspectrum1} shows the neutron energy spectrum in scintillator 
(filled circles) in comparison with the LVD data (open circles with 
error bars) \cite{lvd}. LVD reported in fact the number of events as a 
function of energy deposition, which is not equal to the neutron energy. 
The major difference comes probably from the quenching of protons in 
the scintillator, assuming that the energy deposition is due to the 
energy loss of protons from elastically scattered neutrons. The LVD 
data points, presented in Figure \ref{nspectrum1} have been corrected 
for the quenching factor typical for organic liquid scintillators
\cite{quenching}. The agreement is 
reasonably good taking into account large uncertainties in the 
conversion of visible energy into neutron energy.

\begin{figure}[htb]
\begin{center}
\epsfig{figure=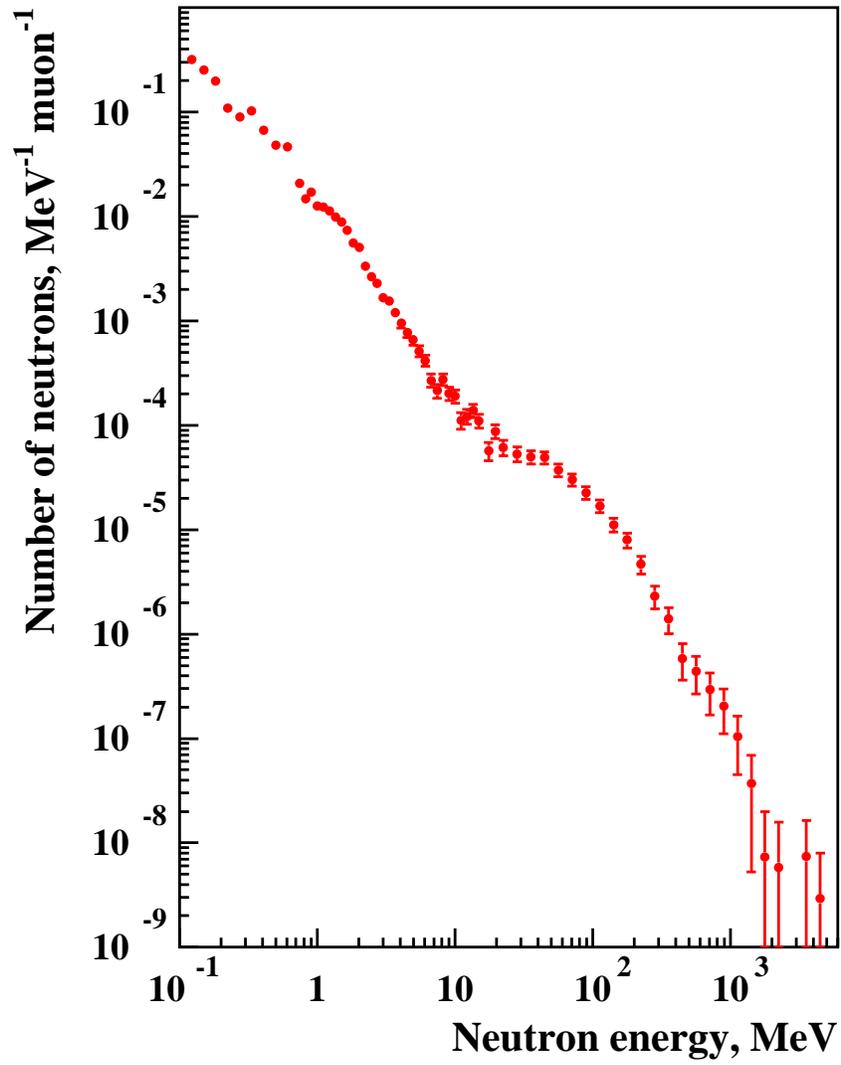,height=15cm}
\caption{Neutron energy spectrum at the boundary between
salt and cavern. }
\label{nspectrum2}
\end{center}
\end{figure}

The next step is calculation of the energy spectrum of neutrons coming
from the rock into the laboratory hall or cavern. We have carried out such 
a simulation for salt with the real muon energy spectrum for Boulby. 
The volume of the salt region was taken as 
$20 \times 20 \times 20$ m$^{3}$, with the cavern 
for the detector of size $6 \times 6 \times 5$ m$^{3}$.
The top of the cavern was placed at a depth of 10 m from the top 
of the salt region. 
The neutrons in the simulations did not stop in the cavern but were
propagated to the opposite wall where they could be scattered back into
the cavern and could be counted for the second time.
Figure \ref{nspectrum2} shows the simulated neutron 
energy spectrum at the salt/cavern boundary. To obtain the 
neutron flux in units MeV$^{-1}$ cm$^{-2}$ s$^{-1}$ 
the differential spectrum plotted on Figure \ref{nspectrum2}
has to be multiplied by the muon flux (see Table \ref{intensity}).
The total number of 
neutrons entering the cavern is about 
$5.8 \times 10^{-10}$ cm$^{-2}$ s$^{-1}$ above 1 MeV
at 3 km w.e. in Boulby rock. 
The flux on an actual detector can
be different from the flux on the boundary salt/cavern due to the
interactions of neutrons in the detector itself.

\begin{figure}[htb]
\begin{center}
\epsfig{figure=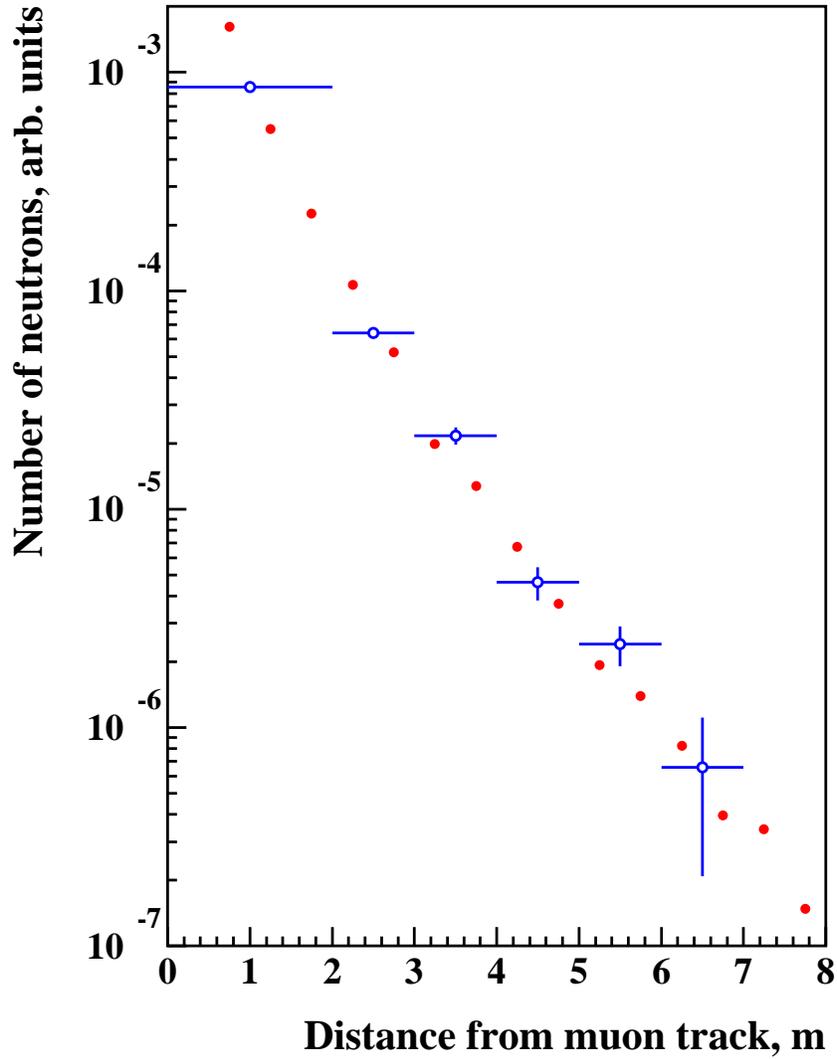,height=15cm}
\caption{Simulated neutron rate as a function of distance from the 
muon track in scintillator (filled circles) 
in comparison with the LVD data \cite{lvd}. 
LVD data are
normalised to the calculated distribution to get better
visual agreement. 
The absolute normalisation is provided by the total neutron
production rate (see Figure \ref{muenergy}). }
\label{nlateral}
\end{center}
\end{figure}

One of the most important features of the muon-induced neutron 
background relevant for dark matter and neutrino experiments is the 
lateral distribution of neutrons, i.e. the number of neutrons as a 
function of distance from muon track. Such a distribution gives the 
probability that a neutron can mimic an expected signal at various 
distances from a muon track, if neither muon or neutron are 
detected by an active veto.
Figure \ref{nlateral} shows the simulated lateral distribution of 
neutrons in scintillator in comparison with the LVD data \cite{lvd}. 
The agreement between them is very good, although LVD is not a 
detector with a single uniform medium, but has a modular structure 
with gaps between modules. 

\vspace{0.5cm}
{\large \bf 4. Conclusions}
\vspace{0.3cm}

We have discussed muon-induced neutron background relevant to dark 
matter and neutrino experiments.
We have presented calculations of muon fluxes and energy spectra at 
various depths underground and estimated their uncertainties.
Neutron production by cosmic-ray muons was simulated for various 
muon energies and various materials. We found reasonably good 
agreement with the recent experimental data of the LVD experiment 
\cite{lvd}. The MUSIC code with associated packages (SIAM and MUSUN),
together with FLUKA, provide a good tool for simulating muon fluxes
and muon-induced neutron background at various depths underground.
Our simulation is the starting point of a 
three-dimensional Monte Carlo to study the neutron
background for any detector.

\vspace{0.5cm}
{\large \bf 5. Acknowledgments}
\vspace{0.3cm}

The authors wish to thank the members of the 
UK Dark Matter Collaboration and the Boulby Dark Matter Collaboration
for valuable discussions. We are grateful, in particular, to
Prof. P. F. Smith, Dr. N. J. T. Smith, Dr. J. D. Lewin and
Prof. G. Gerbier, for useful comments. 
We thank also Dr. I. Dawson for his help to run and understand 
FLUKA.

\pagebreak

\begin{table}[htb]
\caption{Muon intensities at various depths underground for standard 
and Boulby rock and two parameterisations of muon spectrum at the
surface. 
Column 1 -- depth in kilometres of water equivalent, km w.e.; 
column 2 -- vertical muon intensity in standard rock in 
cm$^{-2}$ s$^{-1}$ sr$^{-1}$ with parameterisation of the muon spectrum at 
sea level according to the best fit to LVD data \cite{lvd1}; 
column 3 -- global intensity (integrated 
over solid angle for a spherical detector) in standard rock 
(cm$^{-2}$ s$^{-1}$) for a flat surface with LVD parameterisation 
of the muon spectrum; 
column 4 -- global intensity in standard rock (cm$^{-2}$ s$^{-1}$) 
for a flat surface with Gaisser's parameterisation 
of the muon spectrum at sea level \cite{gaisser}; 
column 5 -- global intensity for Boulby rock ($<Z>=11.7, <A>=23.6$)
with LVD parameterisation of the muon spectrum.}
\label{intensity}
\vspace{1cm}
\begin{center}
\begin{tabular}{|c|c|c|c|c|}\hline
$X$, km w.e. & $I_{\mu}^{vert}$, s.r., LVD & $I_{\mu}$, s.r., LVD & 
$I_{\mu}$, s.r., Gaisser & $I_{\mu}$, Boulby, LVD \\
\hline
0.5 & $1.08 \times 10^{-5}$  & $2.12 \times 10^{-5}$  &
 $1.70 \times 10^{-5}$  & $2.12 \times 10^{-5}$ \\
1.0 & $1.48 \times 10^{-6}$  & $2.60 \times 10^{-6}$  &
 $2.20 \times 10^{-6}$  & $2.57 \times 10^{-6}$ \\
2.0 & $1.39 \times 10^{-7}$  & $2.01 \times 10^{-7}$  &
 $1.81 \times 10^{-7}$  & $1.93 \times 10^{-7}$ \\
3.0 & $2.59 \times 10^{-8}$  & $3.15 \times 10^{-8}$  &
 $2.94 \times 10^{-8}$  & $2.91 \times 10^{-8}$ \\
4.0 & $6.27 \times 10^{-9}$  & $6.54 \times 10^{-9}$  &
 $6.33 \times 10^{-9}$  & $5.84 \times 10^{-9}$ \\
5.0 & $1.74 \times 10^{-9}$  & $1.58 \times 10^{-9}$  &
 $1.58 \times 10^{-9}$  & $1.36 \times 10^{-9}$ \\
6.0 & $5.23 \times 10^{-10}$ & $4.21 \times 10^{-10}$ &
 $4.30 \times 10^{-10}$ & $3.48 \times 10^{-10}$ \\
7.0 & $1.65 \times 10^{-10}$ & $1.18 \times 10^{-10}$ &
 $1.24 \times 10^{-10}$ & $9.41 \times 10^{-11}$ \\
8.0 & $5.38 \times 10^{-11}$ & $3.47 \times 10^{-11}$ &
 $3.73 \times 10^{-11}$ & $2.65 \times 10^{-11}$ \\
9.0 & $1.79 \times 10^{-11}$ & $1.05 \times 10^{-11}$ &
 $1.15 \times 10^{-11}$ & $7.69 \times 10^{-12}$ \\
10.0& $6.06 \times 10^{-12}$ & $3.25 \times 10^{-12}$ &
 $3.65 \times 10^{-12}$ & $2.27 \times 10^{-12}$ \\
\hline
\end{tabular}
\end{center}
\end{table}

\pagebreak

\begin{table}[htb]
\caption{Mean muon energies in GeV at various depths underground 
for standard 
and Boulby rock and two parameterisations of the muon spectrum at 
the surface. 
Column 1 -- depth in kilometres of water equivalent, km w.e.; 
column 2 -- mean muon energy for vertical muon flux in standard rock
with parameterisation of the muon spectrum at 
sea level according to the best fit to the LVD data \cite{lvd1}; 
column 3 --  mean muon energy for global muon flux in standard rock 
for a flat surface with LVD parameterisation of the muon spectrum; 
column 4 --  mean muon energy for global muon flux in standard rock
for a flat surface with Gaisser's parameterisation 
of the muon spectrum at sea level \cite{gaisser}; 
column 5 -- mean muon energy for global muon flux for Boulby rock
with LVD parameterisation of the muon spectrum;
column 6 -- mean muon energy for muon flux at 60$^{o}$ in standard 
rock with LVD parameterisation of the muon spectrum (for this column the 
values in column 1 show the slant depth instead of vertical depth).}
\label{energy}
\vspace{1cm}
\begin{center}
\begin{tabular}{|c|c|c|c|c|c|}\hline
$X$ & $<E_{\mu}^{vert}>$ & $<E_{\mu}>$ & $<E_{\mu}>$ & $<E_{\mu}>$ & 
$<E_{\mu}^{60}>$ \\
km w.e. & s.r., LVD & s.r., LVD & s.r., Gaisser & Boulby, LVD &
s.r., LVD \\
\hline
0.5 &  68 &  92 &  97 &  91 &  74 \\
1.0 & 120 & 150 & 157 & 147 & 127 \\
2.0 & 197 & 226 & 236 & 220 & 205 \\
3.0 & 249 & 273 & 285 & 264 & 256 \\
4.0 & 286 & 304 & 316 & 293 & 292 \\
5.0 & 312 & 324 & 337 & 312 & 316 \\
6.0 & 329 & 338 & 351 & 325 & 332 \\
7.0 & 341 & 348 & 361 & 334 & 343 \\
8.0 & 350 & 356 & 369 & 340 & 351 \\
9.0 & 358 & 361 & 375 & 345 & 357 \\
10.0& 362 & 365 & 380 & 349 & 360 \\
\hline
\end{tabular}
\end{center}
\end{table}

\pagebreak

\pagebreak

\pagebreak

\pagebreak

\pagebreak

\pagebreak

\end{document}